\documentclass[preprint,showpacs,preprintnumbers,amsmath,amssymb,showkeys]{revtex4}

\usepackage{graphicx}
\usepackage{dcolumn}
\usepackage{bm}

\begin{document}

\preprint{Appl. Phys. A}

\title{Wave propagation in the anisotropic metamaterial with single-sheeted hyperboloid dispersion relation}
\author{Hailu Luo}\thanks{
Fax: +86-25-83326028, E-mail: hailuluo@gmail.com}
\author{Zhongzhou Ren}
\author{Weixing Shu}
\author{Fei Li}
\affiliation{ Department of Physics, Nanjing University, Nanjing
210008, China}
\date{\today}

\begin{abstract}
We investigate the wave propagation in the anisotropic
metamaterial with single-sheeted hyperboloid dispersion relation.
Based on boundary conditions and dispersion relations, we find
that the opposite amphoteric refraction, such that E- (or H-)
polarized waves are positively refracted whereas H- (or E-)
polarized waves are negatively refracted, can occur at the
interface associated with the anisotropic metamaterial. Under a
certain condition, both E- and H-polarized waves can exhibit the
same single-sheeted hyperboloid or straight dispersion relation,
while the two polarized waves exhibit different propagation
characteristics. We expect some potential device applications can
be derived based on based on the unique amphoteric refraction
properties.
\end{abstract}

\maketitle

\section{Introduction}\label{Introduction}
In classic electrodynamics, it is well known that the
electrodynamic properties of anisotropic materials are
significantly different from those of isotropic
materials~\cite{Yariv1984,Born1999}. In general, E- and H-
polarized waves propagate in different directions in an
anisotropic material. For regular anisotropic materials, all
tensor elements of permittivity $\boldsymbol{\varepsilon}$ and
permeability $\boldsymbol{\mu}$ should be positive. The advent of
a new class of metamaterials with negative permittivity and
permeability has attained considerable
attention~\cite{Veselago1968,Smith2000,Pendry2000,Shelby2001,Parazzoli2003,Houck2003}.
Lindell {\it et al.} \cite{Lindell2001} have shown that anomalous
negative refraction can occur at an interface associated with an
anisotropic metamaterials, which does not necessarily require that
all tensor elements of $\boldsymbol{\varepsilon}$ and
$\boldsymbol{\mu}$ have negative values. Recently, The studies of
such anisotropic metamaterials have received much interest and
attention~\cite{Hu2002,Smith2003,Zhou2003,Luo2005,Shen2005,Luo2006a,Luo2006b,Depine2006a,Depine2006b}.

The electromagnetic plane waves propagating in the conventional
anisotropic crystal exhibit sphere or ellipsoid wave-vector
surfaces~\cite{Yariv1984,Born1999}. In the anisotropic
metamaterial, however, the electromagnetic plane waves can exhibit
two new kinds of wave-vector surfaces, that is, single-sheeted or
double-sheeted hyperboloid~\cite{Lindell2001,Hu2002,Smith2003}. As
a result, some new wave propagation characteristics will appear.
Because of the importance in achieving some potential
applications, we will focus our interesting on the anisotropic
metamaterial in which both E- and H-polarized waves have the
single-sheeted hyperboloid dispersion relations.

In this paper, we present a detailed investigation on the wave
propagation characteristics in the anisotropic metamaterial with
single-sheeted hyperboloid dispersion relation. First we derive
the single-sheeted hyperboloid dispersion relation from the
Maxwell equations. Next we explore the propagation properties of
wave vector and Poynting vector, and show that E- and H-polarized
waves will exhibit opposite amphoteric (positive or negative)
refraction characteristics. Then we want to explore the case that
both E- and H-polarized have the same single-sheeted hyperboloid
dispersion relation, and find this kind of media can not be
regarded as quasiisotropic. Finally we study the waves propagation
in some special plane, and find both E- and H-polarized waves can
exhibit the same straight dispersion relation.

\section{Plane-wave propagation in anisotropic media}\label{sec2}
It is currently well accepted that a better model is to consider
anisotropic constitutive parameters, which can be diagonalized in
the coordinate system collinear with the principal axes of the
metamaterial. If we take the principal axis as the $z$ axis, the
permittivity and permeability tensors have the following forms:
\begin{eqnarray}
\boldsymbol{\varepsilon}=\left(
\begin{array}{ccc}
\varepsilon_x  &0 &0 \\
0 & \varepsilon_y &0\\
0 &0 & \varepsilon_z
\end{array}
\right), ~~~\boldsymbol{\mu}=\left(
\begin{array}{ccc}
\mu_x &0 &0 \\
0 & \mu_y &0\\
0 &0 & \mu_z
\end{array}
\right).\label{matrix}
\end{eqnarray}
where $\varepsilon_i$ and $\mu_i$ ($i=x,y,z$) are the permittivity
and permeability constants in the principal coordinate system.

Following the standard proceed, we choose the $z$ axis to be
normal to the interface, the $x$ and $y$ axes locate at the plane
of the interface. We consider the propagation of a planar wave
with angular frequency $\omega$ as ${\bf E}={\bf E}_0 e^{i {\bf k}
\cdot {\bf r}-i \omega t}$ and ${\bf H} = {\bf H}_0 e^{i {\bf k}
\cdot {\bf r}-i\omega t}$ from the free space into an anisotropic
metamaterial. The field can be described by Maxwell's equations
\begin{eqnarray}
\nabla\times {\bf E} &=& - \frac{\partial {\bf B}}{\partial
t},~~~{\bf B} =\mu_0 \boldsymbol{\mu}\cdot{\bf H},\nonumber\\
\nabla\times {\bf H} &=&  \frac{\partial {\bf D}}{\partial
t},~~~~~{\bf D} =\varepsilon_0 \boldsymbol{\varepsilon} \cdot {\bf
E}.
\end{eqnarray}

For the plane waves in free space, Maxwell's equations yield the
accompanying dispersion relation has the familiar form:
\begin{equation}
 k_{x}^2+ k_{y}^2+k_{z}^2=\frac{\omega^2}{c^2}. \label{D1}
\end{equation}
Here $k_i$ is the $i$ component of the incident wave vector.
$\omega$ is the frequency, and $c$ is the speed of light in
vacuum. A more careful calculation of Maxwell's equations gives
the dispersion relations in anisotropic media:
\begin{eqnarray}
\left(\frac{ q_{x}^2}{\varepsilon_y
\mu_z}+\frac{q_{y}^2}{\varepsilon_x
\mu_z}+\frac{q_{z}^2}{\varepsilon_y
\mu_x}-\frac{\omega^2}{c^2}\right)\left(\frac{
q_{x}^2}{\varepsilon_z \mu_y}+\frac{q_{y}^2}{\varepsilon_z
\mu_x}+\frac{q_{z}^2}{\varepsilon_x
\mu_y}-\frac{\omega^2}{c^2}\right)=0.\label{D2}
\end{eqnarray}
where $q_{i}$ represents the $i$ component of transmitted
wave-vector. The above equation can be represented by a
three-dimensional surface in wave-vector space. This surface is
known as the normal surface and consists of two shells. It can be
easily shown that there are two types of linearly polarized plane
waves, namely E-polarized and H-polarized plane waves.

To simplify the proceeding analyses, in Eq.~(\ref{D2}) we have
introduced the following relation:
\begin{equation}
\left(\frac{\varepsilon_x }{\mu_x}-\frac{\varepsilon_y
}{\mu_y}\right)\left(\frac{\varepsilon_x
}{\mu_x}-\frac{\varepsilon_z }{\mu_z}\right)=0. \label{UC}
\end{equation}
It should be mentioned that Eq.~(\ref{UC}) is a necessary but not
a sufficient condition for uniaxially anisotropic
media~\cite{Shen2005,Luo2006a}. Without loss of the generality, we
introduce the condition $\varepsilon_x /\mu_x=\varepsilon_y
/\mu_y$. In general, the corresponding wave-vector surfaces are a
combination of ellipsoid or single-sheeted hyperboloid or
double-sheeted hyperboloid. Because of the importance in some
potential applications, we focused our interest on the case that
both E- and H-polarized waves exhibit single-sheeted hyperboloid
wave-vector surfaces.

\section{single-sheeted hyperboloid dispersion relation}
\begin{figure}
\includegraphics[width=8cm]{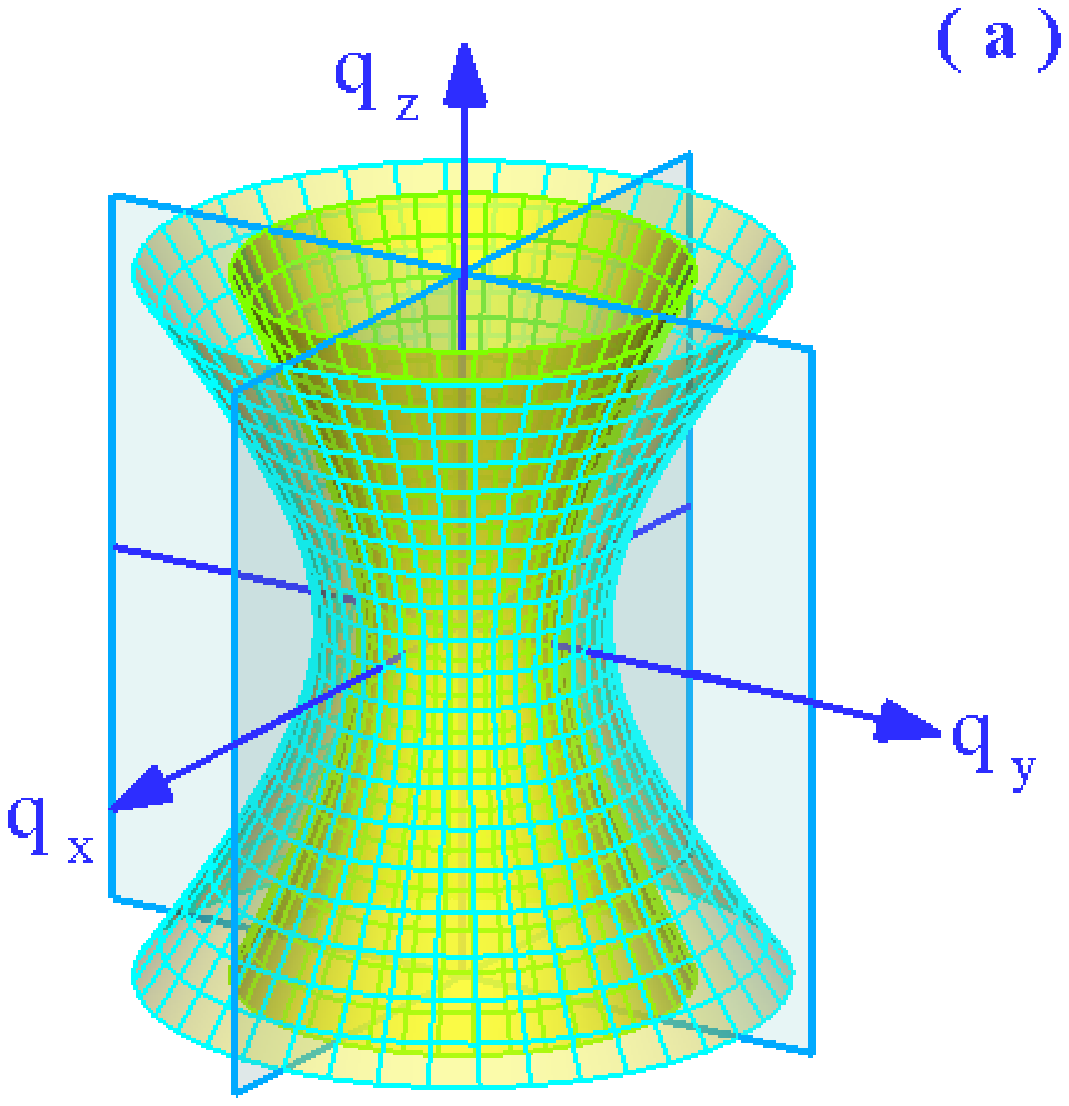}
\includegraphics[width=8cm]{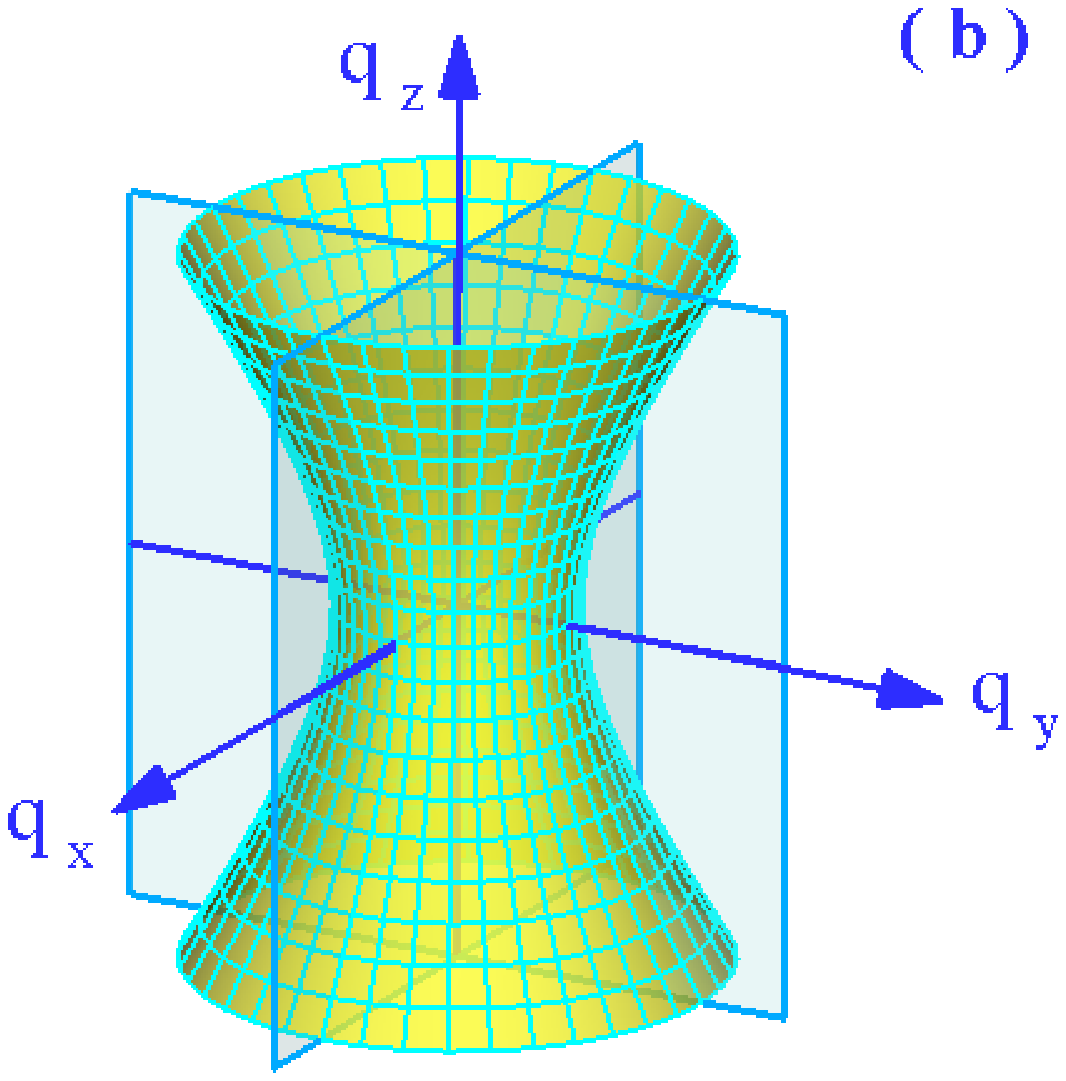}
\caption{\label{Fig1} The two single-sheeted hyperboloid have the
same revolution axis. (a) E- and H-polarized waves have the
different wave-vector surfaces, (b) E- and H-polarized waves
exhibit the same wave-vector surface.}
\end{figure}

In this section, we want to explore the case that the two
single-sheeted hyperboloid wave-vector surface do not intersect.
We assume the revolution axes of the two single-sheeted
hyperboloid coincide with $z$ axis as shown in Fig.~\ref{Fig1}(a).
The anisotropic parameters should satisfy the conditions:
\begin{equation}
\frac{\varepsilon_x}{\mu_x}=\frac{\varepsilon_y }{\mu_y}\neq
\frac{\varepsilon_z }{\mu_z},~~\frac{\varepsilon_i
}{\mu_i}<0,~~\frac{\varepsilon_x
}{\varepsilon_y}>0,~~\frac{\varepsilon_x }{\varepsilon_z}<0.
\end{equation}

First we explore the refraction of the wave vector. The
$z$-component of the wave vector can be found by the solution of
Eq. (\ref{D2}), which yields
\begin{equation}
 q_z^E = \sigma\sqrt {\varepsilon_y \mu_x k_0^2-\varepsilon_y \mu_x \left(\frac{q_{x}^2}{\varepsilon_y
\mu_z}+\frac{q_{y}^2}{\varepsilon_x \mu_z}\right)}, \label{qzh}
\end{equation}
\begin{equation}
 q_z^H = \sigma\sqrt {\varepsilon_x\mu_y k_0^2-\varepsilon_x \mu_y \left(\frac{q_{x}^2}{\varepsilon_z
\mu_y}+\frac{q_{y}^2}{\varepsilon_z \mu_x}\right)}, \label{qze}
\end{equation}
for E- and H-polarized waves, respectively. Here $\sigma=+1$ or
$\sigma=-1$, the choice of the sign ensures that light power
propagates away from the surface to the $+z$ direction. Without
loss of generality, we assume the wave vector locate at the $x-z$
plane ($k_y=q_y=0$). The incident angle of light is given by
\begin{equation}
\theta_I =\tan^{-1}\left[\frac{k_x}{k_{z}}\right].
\end{equation}
The refraction angles of the transmitted wave vector or phase of
E- and H-polarized waves can be written as
\begin{equation}
\beta_P^{E}=\tan^{-1}\left[\frac{q_x^{E}}{q_{z}^{E}}\right],~~~
\beta_P^{H}=\tan^{-1}\left[\frac{q_x^{H}}{q_{z}^{H}}\right].\label{AP}
\end{equation}

Next we  want to the transmission of the Poynting vector. It
should be noted that the actual direction of light is defined by
the time-averaged Poynting vector ${\bf S} =\frac{1}{2} {\bf
Re}({\bf E}^\ast\times \bf{H})$. For E- polarized incident waves,
the transmitted Poynting vector is given by
\begin{equation}
{\bf S}_T^{E}=Re \left[\frac{E_0^2 q_x^{E}}{2 \omega \mu_z}{\bf
e}_x+\frac{E_0^2 q_z^{E}}{2\omega\mu_x}{\bf e}_z\right].\label{SE}
\end{equation}
Analogously, for H-polarized incident waves, the transmitted Poynting vector is
given by
\begin{equation}
{\bf S}_T^{H}=Re\left[\frac{H_0^2 q_x^{H}}{ 2
\omega\varepsilon_z}{\bf e}_x+\frac{H_0^2
q_z^{H}}{2\omega\varepsilon_x}{\bf e}_z\right].\label{SH}
\end{equation}
The refraction angles of Poynting vector of E- and H- polarized
incident waves can be obtained as
\begin{equation}
\beta_S^{E}= \tan^{-1}\left[\frac{S_{Tx}^{E}}{S_{Tz}^{E}}\right],
~~~ \beta_S^{H}=
\tan^{-1}\left[\frac{S_{Tx}^{H}}{S_{Tz}^{H}}\right].\label{AS}
\end{equation}
In free space, the Poynting vector is parallel to the incident
wave vector. Because of the anisotropy, the transmitted Poynting
vector is not necessarily parallel to the refracted wave vector.

From the boundary condition $q_x=k_x$, we can obtain two
possibilities for the refracted wave vector. Energy conservation
requires that the $z$ component of Poynting vector must propagate
away from the interface, for instance, $q_z^{E}/\mu_x>0$ and
$q_z^{H}/\varepsilon_x>0$. Then the signs of $q_z^E$ and $q_z^H$
can be determined easily. The corresponding Poynting vector should
be drawn perpendicularly to the dispersion
contour~\cite{Born1999}. Thus we can obtain two possibilities
(inward or outward), while only the Poynting vector with
$S_{Tz}>0$ is causal.

In general, to distinguish the positive and negative refraction in
the anisotropic metamaterials, we must calculate the direction of
the Poynting vector with respect to the wave vector. Positive
refraction means ${\bf q}_x\cdot{\bf S}_{T}>0$, and negative
refraction means ${\bf q}_x\cdot{\bf S}_{T}<0$~\cite{Luo2006b}.
Evidently, from Eqs.~(\ref{SE}) and (\ref{SH}) we can get
\begin{equation}
{\bf q}_x\cdot{\bf S}_{T}^{E}=\frac{T_E^2 E_0^2 q_x^2}{2 \omega
\mu_z},~~~ {\bf q}_x\cdot{\bf S}_{T}^{H}=\frac{T_H^2 H_0^2
q_x^2}{2 \omega \varepsilon_z}.
\end{equation}
The underlying secret of this anisotropic metamaterial is that
$\varepsilon_z$ and  $\mu_z$ always have the opposite signs.
Clearly, we can easily find that E- and H-polarized waves will
exhibit opposite amphoteric refraction, such that E- (or H-)
polarized waves are positively refracted whereas H- (or E-)
polarized waves are negatively refracted. The opposite amphoteric
refraction is one of the most interesting peculiar properties of
the anisotropic metamaterial.

In the next step, we wish to discuss the interesting anisotropic
metamaterial, which never exist in the conventional anisotropic
crystal. For the special case, if the anisotropic parameters
satisfy the following conditions:
\begin{equation}
\frac{\varepsilon_x}{\mu_x}=\frac{\varepsilon_y
}{\mu_y}=\frac{\varepsilon_z }{\mu_z}<0,~~\frac{\varepsilon_x
}{\varepsilon_y}>0,~~\frac{\varepsilon_x }{\varepsilon_z}<0,
\end{equation}
the wave-vector surfaces of E- and H-polarized waves are the same
single-sheeted hyperbola as shown in Fig.~\ref{Fig1}(b). In this
case
\begin{equation}
\beta_P^{E}=-\beta_P^{H},~~~\beta_S^{E}=-\beta_S^{H}.\label{II1}
\end{equation}
It should be mentioned that in quasiisotropic metamaterial, E- and
H-polarized waves have the same wave-vector surface and exhibit
the same propagation properties~\cite{Luo2006b}. While in this
special case E- and H-polarized waves exhibit the opposite
amphoteric refraction, even if the two polarized waves have the
same wave-vector surface. We thus conclude that this special kind
of anisotropic metamaterial can not be regarded as a
quasiisotropic one.

Because of the importance in potential application, we want to mention the splitting angle
between E- and H-polarized waves. Because the actual direction of light is defined
by the time-averaged Poynting vector,
the splitting angle between E- and H-polarized waves can be defined as
\begin{equation}
\Phi=\beta_S^{E}-\beta_S^{H}.\label{AS}
\end{equation}
The opposite amphoteric refraction suggest that a large splitting
angle can be obtained. The large beam splitting angle and
splitting distance are preferable for practical applications,
especially in the field of optical communication systems.

\section{Straight Lines dispersion relations}\label{sec2}
\begin{figure}
\includegraphics[width=8cm]{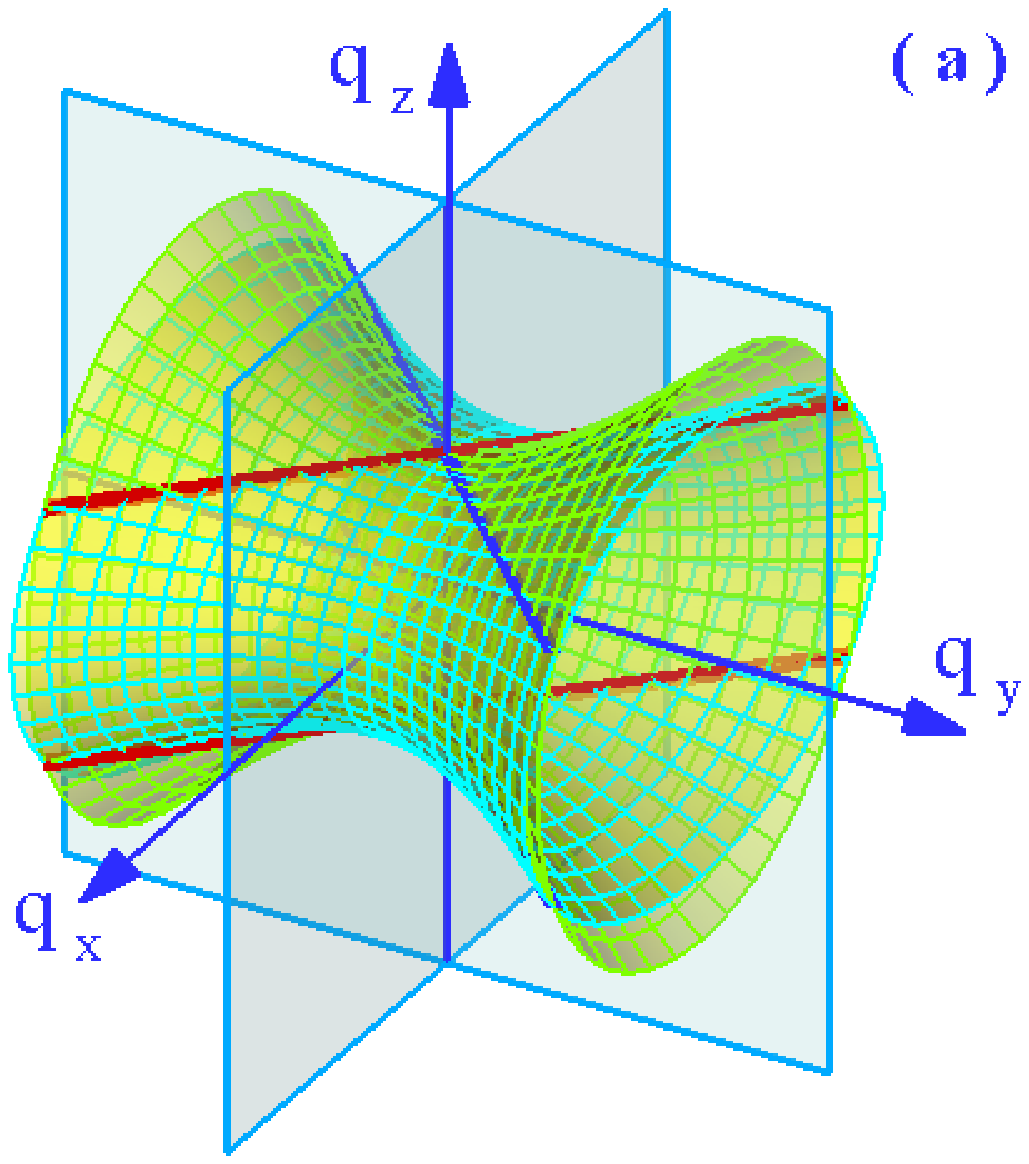}
\includegraphics[width=8cm]{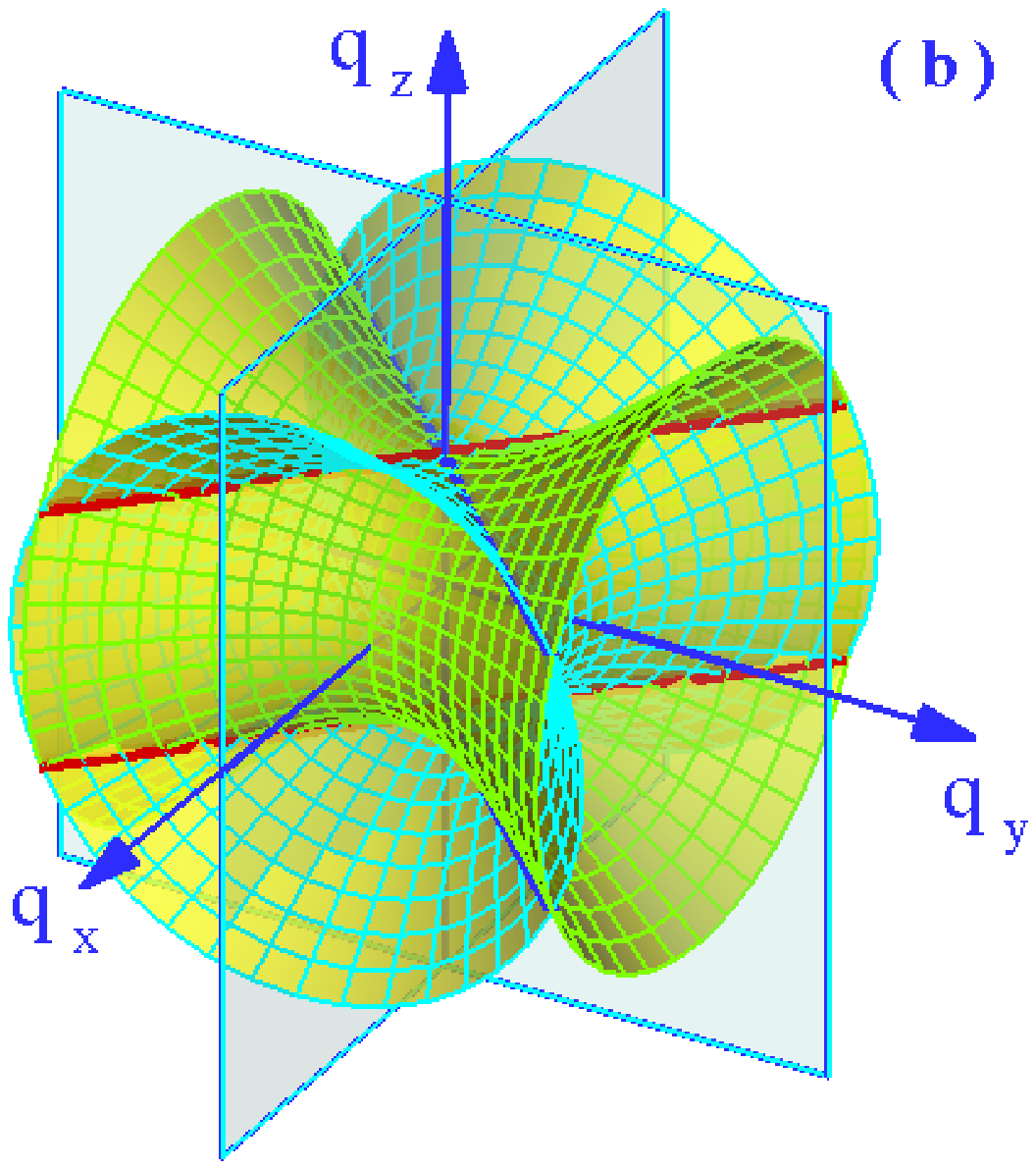}
\caption{\label{Fig2} The two single-sheeted hyperboloid intersect
in four straight lines: (a) The revolution axes of the two
single-sheeted hyperboloid dispersion surface coincide with
direction. (b) The revolution axes of the two single-sheeted
hyperboloid are perpendicular to each other.}
\end{figure}

In this section, we are interested in studying the case that two
single-sheeted hyperboloid intersect each other. There exist two
types which can be formed from combinations
of the material parameter tensor elements.\\
Type I.~ The anisotropic parameters satisfy the conditions:
\begin{equation}
\frac{\varepsilon_x}{\mu_x}=\frac{\varepsilon_y
}{\mu_y}\neq\frac{\varepsilon_z}{\mu_z},~~
\frac{\varepsilon_i}{\mu_i}<0,~~\frac{\varepsilon_x}{\varepsilon_y}<0,~~
\frac{\varepsilon_x}{\varepsilon_z}>0.
\end{equation}
the two single-sheeted hyperboloid have the same revolution axis
as shown in Fig.~\ref{Fig2}(a).
Here we choose the revolution axis coincide with $y$ axis.\\
Type II. The anisotropic parameters satisfy the conditions:
\begin{equation}
\frac{\varepsilon_x}{\mu_x}=\frac{\varepsilon_y }{\mu_y}<0,~~
\frac{\varepsilon_z}{\mu_z}>0,~~
\frac{\varepsilon_x}{\varepsilon_y}<0,~~
\frac{\varepsilon_x}{\varepsilon_z}>0,
\end{equation}
the revolution axes of the two single-sheeted hyperboloid are
perpendicular to each other as depicted in Fig.~\ref{Fig2}(b).

In general, the two wave-vector surfaces intersect in a curve.
Under introducing the condition $\varepsilon_x
/\mu_x=\varepsilon_y /\mu_y$, the two single-sheeted hyperboloid
can intersect in four straight lines
\begin{equation}
\left(\frac{1}{\varepsilon_y\mu_z}-\frac{1 }{\varepsilon_z
\mu_y}\right)q_x^2+\left(\frac{1}{\varepsilon_x\mu_z}-\frac{1
}{\varepsilon_z \mu_x}\right)q_y^2=0. \label{sl}
\end{equation}
Obviously, from Eq.~(\ref{sl}), we can find four straight lines
locate at the plane with the azimuth angles
\begin{equation}
\varphi_c=\pm\arctan\left[\sqrt{\frac{\varepsilon_y\mu_y(
\varepsilon_z \mu_x-\varepsilon_x\mu_z)}{\varepsilon_x
\mu_x(\varepsilon_z \mu_y-\varepsilon_y\mu_z)}}\right]. \label{PA}
\end{equation}
Now, we want to enquires: Whether E- and H-polarized waves exhibit
the same propagation feature or the special plane can be regarded
as a plane axes.

In order to explore the wave propagation in the special plane, we
illustrate the dispersion relation in three dimensions. The
incident wave vector incident with an azimuth angle $\varphi$ from
free space into the anisotropic metamaterial. Then the incident
angle of light is given by
\begin{equation}
\theta_I =\tan^{-1}\left[\frac{k_\perp}{k_{z}}\right],
\end{equation}
where $k_\perp=\sqrt{k_x^2+k_y^2}$, $k_x= k\sin\theta_I
\cos\varphi $, $k_y= k\sin\theta_I \sin\varphi$ and $k=\omega/c$.
The refraction angles of the transmitted wave vector in
anisotropic metamaterial can be found by
\begin{equation}
\beta_P^{E}=
\tan^{-1}\left[\frac{q_\perp^E}{q_z^{E}}\right],~~~\beta_P^{H}=
\tan^{-1}\left[\frac{q_\perp^H}{q_z^{H}}\right],
\end{equation}
where $q_\perp=\sqrt{q_x^2 +q_y^2}$. From the boundary conditions
at the interface $z=0$, the tangential components of the wave
vectors must be continuous, i.e., ${\bf q}_\perp={\bf k}_\perp$.
Based on the dispersion relations and the boundary conditions, the
refraction angle of wave vector can be obtained.

In addition to the refraction of a wave incident at the interface
between free space and the anisotropic metamaterial, we are also
interested in the magnitude of the reflection coefficient, so we
must solve the boundary conditions at the interface. For
E-polarized incident waves, the fields in free space are composed
of the incident and reflected waves, having the form
\begin{equation}
{\bf E} _{I} = [-\sin\varphi {\bf e}_x+\cos\varphi{\bf
e}_y]\exp[i(k_x x+k_y y+k_z z)],\label{EI}
\end{equation}
\begin{equation}
{\bf E} _{R} =[-R_{Ex} \sin\varphi {\bf e}_x+R_{Ey}\cos\varphi{\bf
e}_y] \exp[i(k_x x+k_y y-k_z z)],\label{ER}
\end{equation}
where $R_{Ex}$ and $R_{Ey}$ are the reflection coefficients of the
$x$ and $y$ components, respectively. Matching the boundary
conditions for each $k$ component at $z=0$ gives the complex field
in the anisotropic metamaterial
\begin{equation}
{\bf E} _{T}  = [-T_{Ex} \sin\varphi {\bf e}_x+T_{Ey}
\cos\varphi{\bf e}_y]\exp[i(q_x x+ q_y y+ q_z z)],\label{ET}
\end{equation}
where $T_{Ex}$ and $T_{Ey}$ are the transmission coefficients of
the $x$ and $y$ components, respectively. Based on the boundary
conditions yield the equations
\begin{equation}
1+R_{Ex}=T_{Ex},~~1+R_{Ey}=T_{Ey}.\label{EBXY}
\end{equation}
We next require continuity of the transverse component of the
magnetic field, which can be found from the electric field by use
of the Maxwell curl equation combined with the general
constitutive relation
\begin{equation}
{\bf H} =i \frac{c}{\omega}\boldsymbol{\mu}^{-1}\cdot\nabla\times{\bf
E},
\end{equation}
Equating the $x$ and $y$ components of the magnetic vectors
corresponding to the incident, reflected, and transmitted fields,
we have
\begin{equation}
\mu_y k_z(1-R_{Ex} )=T_{Ex} q_z^{E},~~\mu_x k_z(1-R_{Ey})= T_{Ey}
q_z^{E}.\label{BBXY}
\end{equation}
For E-polarized incident waves, combining Eq.~(\ref{BBXY}) with
Eq.~(\ref{EBXY}), we can obtain the following expressions for the
reflection and transmission coefficients for $x$ and $y$
components
\begin{equation}
R_{Ex}=\frac{\mu_y k_z- q_z^{E}}{\mu_y  k_z+ q_z^{E}},~~~T_{Ex} =
\frac{2 \mu_y k_z}{\mu_y k_z+ q_z^{E}},
\end{equation}
\begin{equation}
R_{Ey}=\frac{\mu_x k_z- q_z^{E}}{\mu_x  k_z+ q_z^{E}},~~~T_{Ey} =
\frac{2 \mu_x k_z}{\mu_x  k_z+ q_z^{E}}.
\end{equation}
Similarly, the reflection and transmission coefficients of
H-polarized waves can be obtained as
\begin{equation}
R_{Hx}=\frac{\varepsilon_y k_z-q_z^{H}}{\varepsilon_y k_z+
q_z^{H}},~~~T_{Hx} = \frac{2 \varepsilon_y k_z}{\varepsilon_y k_z+
q_z^{H}},
\end{equation}
\begin{equation}
R_{Hy}=\frac{\varepsilon_x k_z- q_z^{H}}{\varepsilon_x
k_z+q_z^{H}},~~~T_{Hy} = \frac{2 \varepsilon_x k_z}{\varepsilon_x
k_z+ q_z^{H}}.
\end{equation}

After the reflection and transmission coefficients are determined,
the energy current density ${\bf S}_T$  of the refracted waves can
be obtained. From the boundary condition $q_\perp=k_\perp$, we can
obtain two possibilities for the refracted wave vector. Energy
conservation requires that the $z$ component of transmitted
Poynting vector must propagate away from the interface, for
instance:
\begin{equation}
\frac{T_{Ex}^2\sin^2\varphi q_z^{E} }{2\mu_y} +\frac{T_{Ey}^2
\cos^2\varphi q_z^{E} }{2\mu_x}>0,\label{SEZ}
\end{equation}
\begin{equation}
\frac{T_{Hx}^2\sin^2\varphi q_z^{H} }{2\varepsilon_y}
+\frac{T_{Hy}^2 \cos^2\varphi q_z^{H}
}{2\varepsilon_x}>0.\label{SHZ}
\end{equation}
Then the signs of $q_z^E$ and $q_z^H$  can be determined easily.
The corresponding Poynting vectors should be drawn perpendicularly
to the dispersion contour. Since
$\varepsilon_x/\mu_x=\varepsilon_y /\mu_y<0$, $q_z^{E}$ and
$q_z^{H}$ always have the opposite signs as can be seen from
Eqs.~(\ref{SEZ}) and (\ref{SHZ}). Therefore  E- and H-polarized
waves exhibit different values of wave vectors, even if  the two
polarized waves have the same dispersion relation.

Finally we want to discuss wave propagation in the special plane
where the azimuth angle $\varphi=\varphi_c$. In this special
plane, both E- and H-polarized exhibit the same straight lines
dispersion relation as shown in Fig.~\ref{Fig2}. For propagation
in the direction of the optic axes, there is only one value of
wave vector and, consequently, only one phase velocity. There are,
however, two independent directions of
polarization~\cite{Yariv1984,Born1999}. Clearly, there do not
exist the direction where E- and H-polarized waves will exhibit
the same wave vector (or phase velocity). We thus conclude that
this special plane can not be regarded as a plane axes, if we
think the conventional concept of optical axis is still correct.

\section{Conclusion }\label{sec4}
In conclusion, we have investigated the properties of wave
propagation in the anisotropic metamaterial with single-sheeted
hyperboloid dispersion relation. At the interface associated with
such anisotropic metamaterial, we found E- and H-polarized waves
exhibit opposite amphoteric refraction characteristics, such that
one polarized waves are positively refracted whereas the other
polarized waves are negatively refracted. We have explored the
wave propagation in the unique anisotropic metamaterial, in which
both E- and H-polarized waves have the same single-sheeted
hyperboloid or straight lines dispersion relation. We expect many
potential device applications can be fabricated based on the
special properties of waves propagation discussed above. They can,
for example, be used to construct very efficient
polarization-independent isolators, optical switches, and
polarization splitters.

\begin{acknowledgements}
H. Luo and W. Shu are sincerely grateful to Professor A. Lakhtakia
for many fruitful discussions. This work was supported by projects
of the National Natural Science Foundation of China (Nos. 10125521
and 10535010) and the 973 National Major State Basic Development
of China (No. G2000077400).
\end{acknowledgements}


\begin{references}
\bibitem{Yariv1984} A. Yariv, P. Yeh: \emph{Optical Waves in Crystals}, (John
Wiley and Sons,  New York, 1984)

\bibitem{Born1999} M. Born, E. Wolf: \emph{Principles of Optics}, (Cambridge,
New York, 1999)

\bibitem{Veselago1968}  V.G.  Veselago: Sov. Phys. Usp.  {\bf 10}, 509 (1968).

\bibitem{Smith2000}  D.R. Smith,  W.J. Padilla, D.C. Vier, S.C. Nemat-Nasser, S. Schultz: Phys. Rev. Lett.  {\bf 84}, 4184 (2000).

\bibitem{Pendry2000} J.B. Pendry: Phys. Rev. Lett. {\bf 85}, 3966 (2000).

\bibitem{Shelby2001} R.A. Shelby: D.R. Smith, S. Schultz, Science  {\bf 292}, 77 (2001).

\bibitem{Parazzoli2003}  C.G. Parazzoli, R.B. Greegor, K. Li, B.E C.
Koltenba, M. Tanielian: Phys. Rev. Lett.  {\bf 90}, 107401 (2003).

\bibitem{Houck2003} A.A. Houck, J.B. Brock, I.L. Chuang: Phys. Rev. Lett.  {\bf 90}, 37401 (2003).

\bibitem{Lindell2001} I.V. Lindell, S.A. Tretyakov, K.I.
Nikoskinen, S. Ilvonen, Microw. Opt. Technol. Lett.  {\bf 31}, 129
(2001).

\bibitem{Hu2002}   L. Hu, S.T. Chui: Phys. Rev. B {\bf 66}, 085108 (2002).

\bibitem{Smith2003}  D.R. Smith, D. Schurig:  Phys. Rev. Lett.  {\bf 90}, 077405 (2003).

\bibitem{Zhou2003}   L. Zhou, C.T. Chan, P. Sheng: Phys. Rev. B  {\bf 68} 115424 (2003).

\bibitem{Luo2005} H. Luo, W. Hu, X. Yi, H. Liu, J. Zhu: Opt. Commun.  {\bf 254}, 353 (2005).

\bibitem{Shen2005} N.H. Shen, Q. Wang, J. Chen, Y.X. Fan, J. Ding, H.T. Wang, Y. Tian,
N.B. Ming: Phys. Rev. B  {\bf 72}, 153104 (2005).

\bibitem{Luo2006a} H. Luo, W. Shu, F. Li, Z. Ren: Opt. Commun.
{\bf 267}, 271 (2006).

\bibitem{Luo2006b} H. Luo, W. Hu, W. Shu, F. Li, Z. Ren:
Europhysics Letters {\bf 74}, 1081 (2006).

\bibitem{Depine2006a} R.A. Depine, M.E. Inchaussandague, A. Lakhtakia:
J. Opt. Soc. Am. A {\bf 23}, 949 (2006).

\bibitem{Depine2006b} R.A. Depine, M.E. Inchaussandague, A.
Lakhtakia: J. Opt. Soc. Am. B {\bf 23}, 514 (2006).


\end{references}
\end{document}